\documentclass{article}

\usepackage{spconf,amsmath,makecell} %
\usepackage{cite,mathtools,amssymb,bm}
\usepackage{balance,booktabs,arydshln,multirow}

\usepackage[redeflists]{IEEEtrantools}

\title{Capturing Multi-Resolution Context by Dilated Self-Attention}
\name{Niko Moritz, Takaaki Hori, Jonathan Le Roux}
\address{Mitsubishi Electric Research Laboratories (MERL), Cambridge, MA, USA}

\begin{document}

\bstctlcite{IEEEexample:BSTcontrol}

\ninept
\setlength{\abovedisplayskip}{4pt}
\setlength{\belowdisplayskip}{4pt}
\setlength{\abovecaptionskip}{3pt}
\setlength{\belowcaptionskip}{-3pt}
\setlength{\textfloatsep}{5pt}
\setlength{\parindent}{1em}

\maketitle
\begin{abstract} %

Self-attention has become an important and widely used neural network component that helped to establish new state-of-the-art results for various applications, such as machine translation and automatic speech recognition (ASR). However, the computational complexity of self-attention grows quadratically with the input sequence length. This can be particularly problematic for applications such as ASR, where an input sequence generated from an utterance can be relatively long. In this work, we propose a combination of restricted self-attention and a dilation mechanism, which we refer to as dilated self-attention. The restricted self-attention allows attention to neighboring frames of the query at a high resolution, and the dilation mechanism summarizes distant information to allow attending to it with a lower resolution. Different methods for summarizing distant frames are studied, such as subsampling, mean-pooling, and attention-based pooling. ASR results demonstrate substantial improvements compared to restricted self-attention alone, achieving similar results compared to full-sequence based self-attention with a fraction of the computational costs.
  
\end{abstract}

\vspace{-0.1cm}
\section{Introduction}
\vspace{-0.1cm}

The attention mechanism has become a central component in many neural network architectures for machine translation, speech processing, language modeling, and computer vision \cite{ChorowskiBSCB15,BahdanauCB14,Ramachandran19attInCV,zhao20exploringSA,IrieZSN19}.
It is frequently used as a relay between encoder and decoder neural networks or as a substitute for recurrent neural networks (RNNs) and convolutional neural networks (CNNs) \cite{ParikhTDU16,PoveyHGLK18}.
Recently, the transformer architecture \cite{VaswaniSPUJGKP17}, which utilizes attention throughout all model components, has set new state-of-the-art results for many different machine learning applications, including many speech and audio processing tasks \cite{DongXX18,KaritaCH19,MoritzHLR20,MoritzHLR20b}. 
As attention-based architectures have been successfully applied for various domains, the number of model parameters have tended to increase to further improve results using deeper and wider architectures \cite{shoeybi2019megatronlm,Devlin2019BERTPO}.
However, in particular at inference time, computational costs must be kept small to enable scalability of a model for cloud-based services that serve many users simultaneously or to allow local computation on mobile devices.

The attention mechanism is a method to query information from an input sequence, which can also be regarded as reading from a memory \cite{ChorowskiBSCB15}. In self-attention, each frame of an input sequence is used once as a query to read information from itself \cite{ParikhTDU16,VaswaniSPUJGKP17}.
In automatic speech recognition (ASR), neighboring frames of such a query frame may belong to the same phone, syllable, or word, where detailed information is required to recognize their coherency. On the other hand, distant information is relevant to recognize the context of sounds and words in an utterance and to adapt to speaker or recording characteristics, which typically requires less fine-grained information. 
This line of arguments is also applicable to other tasks, e.g., to machine translation or language modeling, where close-by words are more likely to have a dependent relationship, while only a few distant words or word groups are relevant to trace the semantic context and syntax of a sentence \cite{KhandelwalHQJ18}.

This hypothesis is investigated in this work by combining restricted (or time-restricted) self-attention with a dilation mechanism, whereby a high self-attention resolution for neighboring frames and a lower self-attention resolution for distant information are achieved. The proposed method, named dilated self-attention in analogy to dilated convolution \cite{YuK16_dilatedcnn}, alleviates the quadratic computational cost growth of self-attention with the input sequence length.
Various frame rate reduction methods are studied for the dilation mechanism, including subsampling as well as pooling methods that extract or compress the relevant information within a chunk of frames.
We compare mean-pooling to a newly proposed attention-based pooling approach in this work.
In such a setup, the computational cost of the restricted self-attention grows only linearly with the input sequence length, and the computational costs for attending to a dilation sequence are smaller by a factor $M$ compared to standard self-attention, where $M$ denotes the subsampling or the chunk size of the pooling operation. Thus, the overall complexity of dilated self-attention is significantly reduced, while the full context of an input sequence is still captured with different resolutions.

ASR results are reported for two different data sets using a transformer-based end-to-end ASR system.
It is shown that dilated self-attention can reduce self-attention costs by several orders with almost no performance degradation for offline and streaming ASR applications. In addition, dilated self-attention demonstrates clear advances in terms of word error rates (WERs) over restricted self-attention alone.
An attention-based pooling method is proposed that uses learned query vectors to compute the weighted average of each chunk by attention, where optionally a post-processing stage can be applied, e.g., to merge outputs of multiple attention heads. Among the tested pooling methods, attention-based pooling with post-processing is shown to achieve the highest robustness.

\vspace{-0.2cm}
\section{System Architecture}
\label{sec:architecture}
\vspace{-0.2cm}

In this work, a joint connectionist temporal classification (CTC) and transformer-based end-to-end ASR system is used, which combines the advantages of both model types for training and decoding \cite{WatanabeHKHH17,KaritaCH19} achieving state-of-the-art ASR results \cite{KaritaYWD19} and enabling streaming recognition of encoder-decoder based ASR systems \cite{MoritzHLR20,MoritzHR19c}.

The transformer model leverages two different attention types: encoder-decoder attention and self-attention \cite{VaswaniSPUJGKP17}.
Encoder-decoder attention uses a decoder state as a query for attending to an input sequence, the encoder output.
In self-attention, the queries are computed from the same input sequence, which results in an output sequence of the same length. Both attention types of the transformer model are based on the scaled dot-product attention mechanism,
\begin{equation}
    \text{Attention}(Q, K, V) = \text{Softmax}\left(\dfrac{Q K^\mathsf{T}}{\sqrt{d_k}}\right)V, \label{eq:mhatt} 
\end{equation}
where $Q \in \mathbb{R}^{n_q \times d_q}$, $K \in \mathbb{R}^{n_k \times d_k}$, and $V \in \mathbb{R}^{n_v \times d_v}$ are the queries, keys, and values, where the $d_*$ denote dimensions and the $n_*$ denote sequence lengths, $d_q=d_k$, and $n_k=n_v$ \cite{VaswaniSPUJGKP17}. Instead of using a single attention head, multiple attention heads are used with %
\begin{align}
    \text{MHA}(\hat Q, \hat K, \hat V) &= \text{Concat}_\text{f}(\text{Head}_1, \dots, \text{Head}_{d_h}) W^\text{H} \label{eq:mha}\\
    \text{and } \text{Head}_i &= \text{Attention}(\hat Q W_i^Q, \hat K W_i^K, \hat V W_i^V),  \label{eq:att_heads}
\end{align}
where $\hat Q$, $\hat K$, and $\hat V$ are inputs to the multi-head attention (MHA) layer, $\text{Head}_i$ represents the output of the $i$-th attention head for a total number of $d_h$ heads, $W_i^Q \in \mathbb{R}^{d_\mathrm{model} \times d_q}$, $W_i^K \in \mathbb{R}^{d_\mathrm{model} \times d_k}$, $W_i^V \in \mathbb{R}^{d_\mathrm{model} \times d_v}$ as well as $W^H \in \mathbb{R}^{d_hd_v \times d_\mathrm{model}}$ are trainable weight matrices with typically $d_k=d_v=d_\mathrm{model}/d_h$, and $\text{Concat}_\text{f}$ denotes concatenation along the feature dimension of size $d_v$.

The transformer encoder architecture consists of a two-layer CNN module \textsc{EncCNN} and a stack of self-attention based layers \textsc{EncSA}:
\begin{align}
    X_0 &= \textsc{EncCNN} (X) + \textsc{PE}, \\
    X_{E} &= \textsc{EncSA} (X_0), \label{eq:enc_sa}
\end{align}
where $\textsc{PE}$ are sinusoidal positional encodings \cite{VaswaniSPUJGKP17} and $X$ %
denotes a sequence of acoustic input features, which are 80-dimensional log-mel spectral energies plus 3 extra features for pitch information \cite{HoriWZC17}. Both CNN layers of \textsc{EncCNN} use a stride of size $2$, a kernel size of $3 \times 3$, and a ReLU activation function, which reduces the output frame rate by a factor of 4.
The \textsc{EncSA} module of (\ref{eq:enc_sa}) consists of $E$ layers, where the $e$-th layer, for $e=1,\dots,E$, is a composite of a multi-head self-attention layer
\begin{equation} \label{eq:enc_selfatt}
    X'_e = X_{e-1} + \text{MHA}_{e}(X_{e-1}, X_{e-1}, X_{e-1}),
\end{equation}
and two feed-forward neural networks of inner dimension $d_\mathrm{ff}$ and outer dimension $d_\mathrm{model}$ that are separated by a ReLU activation function as follows:
\begin{align}
    X_{e} &= X'_e + \text{FF}_e(X'_e), \\
    \text{with } \text{FF}_e(X'_e) &= \text{ReLU} (X'_e W_{e,1}^\mathrm{ff} + b_{e,1}^\mathrm{ff}) W_{e,2}^\mathrm{ff} + b_{e,2}^\mathrm{ff},
    \label{eq:FF}
\end{align}
where $W_{e,1}^\mathrm{ff} \in \mathbb{R}^{d_\mathrm{model} \times d_\mathrm{ff}}$, $W_{e,2}^\mathrm{ff} \in \mathbb{R}^{d_\mathrm{ff} \times d_\mathrm{model}}$, $b_{e,1}^\mathrm{ff} \in \mathbb{R}^{d_\mathrm{ff}}$, and $b_{e,2}^\mathrm{ff} \in \mathbb{R}^{d_\mathrm{model}}$ are trainable weight matrices and bias vectors. 

{\allowdisplaybreaks
The transformer objective function is defined as
\begin{equation}
  p_{\text{att}}(Y|X_E) = \prod_{l=1}^{L} p(y_l | \bm y_{1:l-1}, X_E) \label{eq:transformer_objf}
\end{equation}
with label sequence $Y=(y_1,\dots,y_L)$, label subsequence $\bm y_{1:l-1}=(y_1,\dots,y_{l-1})$, and the encoder output sequence $X_E=\allowbreak(\bm x_1^{E},\dots \allowbreak,\bm x_N^{E})$.
The term $p(y_l | \bm y_{1:l-1}, X_E)$ represents the transformer decoder model, which consists of a stack of $D$ layers each applying two MHA layers, one performing self-attention using $\bm y_{1:l-1}$ as an input and one for encoder-decoder attention, followed by a feed-forward neural network similar to (\ref{eq:FF}). Finally, the posterior probabilities are estimated by applying a fully-connected neural network to the last decoder layer and a softmax distribution over that output.

}

The transformer model is trained jointly with the CTC objective function $p_\text{ctc}$ \cite{GravesFGS06,KaritaYWD19} using the multi-objective loss function
\begin{equation}
  \mathcal{L} = -\gamma \log p_\text{ctc} - (1-\gamma) \log p_\text{att},
  \label{eq:loss}
\end{equation}
where hyperparameter $\gamma$ controls the weighting between the two objective functions $p_\text{ctc}$ and $p_\text{att}$.

\begin{figure}[t]
  \centering
  \centerline{\includegraphics[width=\linewidth]{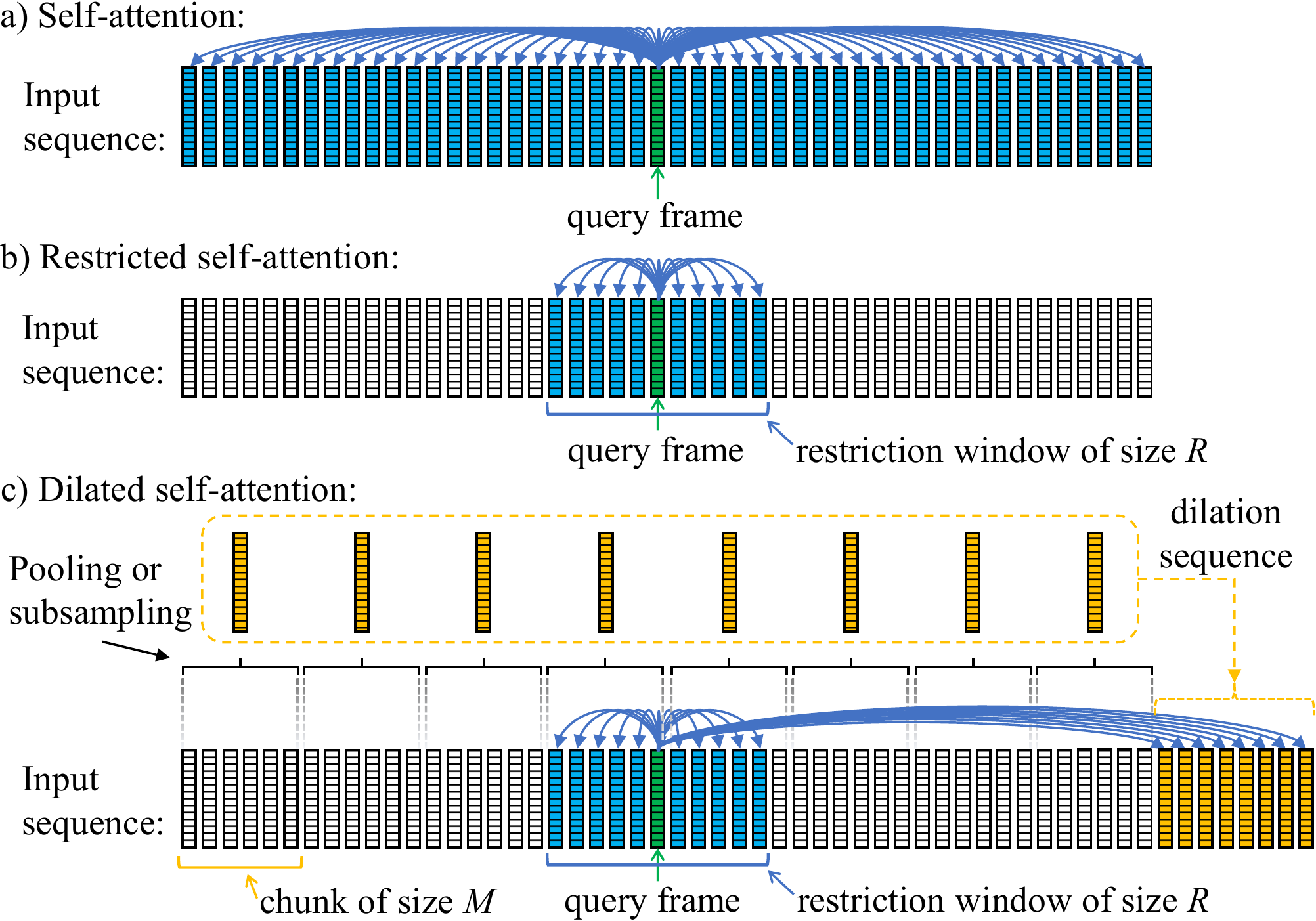}}\vspace{-3mm}
  \caption{Full-sequence based self-attention is shown in a), where each bar of the input sequence represents an input frame or vector. A current query frame is highlighted in green color and blue arrows represent allowed attention connections. Restricted self-attention is depicted in b), where frames within the restriction window are highlighted in blue color. Dilated self-attention is shown in c), where a dilation sequence (in yellow) is generated for the keys and values by a pooling or subsampling mechanism and appended to keys and values of the restricted sequence prior to self-attention. }\vspace{.1cm}
\label{fig:self-att-types}
\end{figure}

\vspace{-0.1cm}
\section{Dilated Self-Attention}
\vspace{-0.1cm}

Self-attention, restricted self-attention, and dilated self-attention are illustrated in Fig.~\ref{fig:self-att-types}. For restricted self-attention, the past and future context (blue frames) relative to a current query vector (green frame) is limited by using a fixed number of look-back and look-ahead frames.
For dilated self-attention, restricted self-attention is combined with a dilation mechanism, which is shown in Fig.~\ref{fig:self-att-types}c. The dilation mechanism subsamples or summarizes the keys/values and appends the generated dilation sequence, which is of lower frame rate than the input sequence, to the restricted keys/values for the restricted self-attention process.

\vspace{-0.1cm}
\subsection{Dilation Mechanisms}
\vspace{-0.1cm}

The dilation mechanism at the $e$-th encoder layer first splits the keys $K_i=X_{e-1}W^K_i=(\bm k^i_1,\dots,\bm k^i_N)$ and values $V_i=X_{e-1}W^V_i=(\bm v^i_1,\dots,\bm v^i_N)$ each of length $N$, cf.\ (\ref{eq:att_heads}), into $L=\lceil \frac{N}{M} \rceil$ non-overlapping key chunks $C^K_{i,l}$ and value chunks $C^V_{i,l}$ each of length $M$, such that
\begin{align}
    C^V_{i,l} &= (\bm v^i_{M(l-1)+1}, \dots, \bm v^i_{M(l-1)+M}), \\
    C^K_{i,l} &= (\bm k^i_{M(l-1)+1}, \dots, \bm k^i_{M(l-1)+M}),
\end{align}
for $l=1,\dots,L$, where the last chunks, $C_{i,L}^V$ and $C_{i,L}^K$, are zero-padded if they have fewer than $M$ frames.
Next, subsampling or pooling techniques are applied to each chunk, to generate dilation sequences $\Delta_i^K$ and $\Delta_i^V$ that are appended to the restricted keys and values, respectively, by modifying (\ref{eq:att_heads}) as follows:
\begin{align} \label{eq:dilated_selfatt}
    \text{Head}_{i,n,e} &= \text{Attention}(\bm x_n^{e-1} W_i^Q, \bar K_{i,n,e}, \bar V_{i,n,e}) \\
    \text{with } \bar K_{i,n,e} &= \text{Concat}_\text{t}(\bm k^i_{n-\nu_\text{lb}:n+\nu_\text{la}}, \Delta_i^K), \\
    \text{and } \bar V_{i,n,e} &= \text{Concat}_\text{t}(\bm v^i_{n-\nu_\text{lb}:n+\nu_\text{la}}, \Delta_i^V),
\end{align}
for $n=1,\dots,N$, where $\nu_\text{lb}$ and $\nu_\text{la}$ denote the number of look-back and look-ahead frames for the time-restriction, which corresponds to a window size of $R=\nu_\text{lb}+\nu_\text{la}+1$, and $\text{Concat}_\text{t}$ denotes concatenation along the time dimension (frames).

The \textbf{subsampling}-based dilation mechanism selects the first frame of each chunk to form the dilation sequences $\Delta_i^K = (\bm k^i_{1},\allowbreak \dots,\bm k^i_{M(l-1)+1},\allowbreak \dots,\allowbreak \bm k^i_{M(L-1)+1})$ and $\Delta_i^V = (\bm v^i_{1},\allowbreak \dots,\bm v^i_{M(l-1)+1},\allowbreak \dots,\allowbreak \bm v^i_{M(L-1)+1})$.
Alternatively to subsampling, pooling methods can be applied to summarize the frames of each chunk. In this work, we compare three different pooling mechanisms: 1) mean-pooling (MP), 2) attention-based pooling (AP), and 3) attention-based pooling with post-processing (AP+PP).

For the \textbf{mean-pooling} (MP) based dilation mechanism, frames in each chunk are averaged to the mean vectors
\begin{align}
    \bm \mu^{[V,K]}_{i,l} &= \frac{1}{M} \sum_{m} C^{[V,K]}_{i,l}[m], %
\end{align}
for $l=1,\dots,L$, where the notation $^{[V,K]}$ denotes the processing of either the values or the keys. We will continue to use this notation for the following equations. The sequence of mean vectors is used to form the dilation sequences $\Delta_i^{[V,K]} = (\bm \mu^{[V,K]}_{i,1},\dots,\bm \mu^{[V,K]}_{i,L})$.

In \textbf{attention-based pooling} (AP), we train embedding vectors to query summary information from the key and value chunks by using the attention mechanism as follows:
\begin{align}
    \bm g^{[V,K]}_{i,l} &= \frac{1}{B} \sum_{b=1}^B \bm a^{[V,K]}_{i,b,l}, \\ \label{eg:att_pool_V}
    \bm a^{[V,K]}_{i,b,l} &= \text{Attention}(\bar{\bm{q}}^{K}_{b}, C^{K}_{i,l}, C^{[V,K]}_{i,l}), \\
    \bar{\bm{q}}^{K}_b &= \text{Embed}^{K}(b),
\end{align}
for $l=1,\dots,L$, where $\bar{\bm{q}}^{K}_b$ %
represents a query, $\text{Embed}^{K}(b)$ maps the attention head numbers $b=1,\dots,B$ to trainable vectors of dimension $d_\text{k}$, and $B$ denotes the total number of attention heads.
The attention outputs $\bm a^{[V,K]}_{i,b,l}$ are averaged along $b=1,\dots,B$ to form the dilation sequences $\Delta_i^{[V,K]}=(\bm g^{[V,K]}_{i,1},\dots,\bm g^{[V,K]}_{i,L})$. 

\textbf{Post-processing} (PP) can be applied to $\bm a^{[V,K]}_{i,l}$ to further process the AP output and to effectively join the output of multiple attention heads using a two-layer feed-forward neural network of inner dimension $d_\text{in}$ and outer dimension $d_\text{[v,k]}$:
\begin{align}
    \bm p^{[V,K]}_{i,l} &= FF^{[V,K]}(\bm a^{[V,K]}_{i,l})+ \bm g^{[V,K]}_{i,l}, \label{eq:appp} \\
    FF^{[V,K]}\!(\bm a^{[V,K]}_{i,l})\! &= \text{ReLU}(\bar{\bm{a}}^{[V,K]}_{i,l} W_1^{[V,K]} \!\!+\! \bm b_1^{[V,K]})W_2^{[V,K]} \!\!+\! \bm b_2^{[V,K]}\!, \\
    \bar{\bm{a}}^{[V,K]}_{i,l} &= \text{Concat}_\text{f}(\bm a^{[V,K]}_{i,1,l},\dots,\bm a^{[V,K]}_{i,B,l}),
\end{align}
where $W_1^{[V,K]} \in \mathbb{R}^{d_\text{[v,k]}B \times d_\text{in}}$, $W_2^{[V,K]} \in \mathbb{R}^{d_\text{in} \times d_\text{[v,k]}}$, $\bm b_1^{[V,K]} \in \mathbb{R}^{d_\text{in}}$, and $\bm b_2^{[V,K]} \in \mathbb{R}^{d_\text{[v,k]}}$ are trainable weight matrices and bias vectors and $\text{Concat}_\text{f}$ denotes concatenation of the vectors $\bm a^{[V,K]}_{i,b,l}$ for $b=1,\dots,B$ along the feature dimension.
The PP results $\bm p_{i,l}^{[V,K]}$ are then used to form the dilation sequences $\Delta_i^{[V,K]}=(\bm p^{[V,K]}_{i,1},\dots,\bm p^{[V,K]}_{i,L})$.

\vspace{-0.1cm}
\subsection{Computational complexity estimation}
\vspace{-0.1cm}

The computational complexity estimation in this work is based on the number of floating-point multiplications of vector and matrix products, which is described here by the $\mathcal{M}$ notation. For simplicity, we ignore in the estimation scalar multiplications as well as additions, since including these operations does not significantly change the relative complexities when comparing the different methods. The complexity of the full-sequence based self-attention process is $\mathcal{M}(N^2 d_\text{model})$, where $N$ denotes the length of an input sequence and $d_\text{model}$ the attention model dimension, cf.\ Section~\ref{sec:architecture}.
The complexity for restricted self-attention is $\mathcal{M}(N R d_\text{model})$, where $R$ is the size of the restriction window, which is constant and typically significantly smaller than $N$. %
The computational complexity of dilated self-attention is $\mathcal{M}(N (R + \lceil \frac{N}{M} \rceil) d_\text{model}) + \xi$, which includes the attention costs for restricted self-attention with the appended dilation sequence plus $\xi$, the complexity of the dilation mechanism.

The complexity $\xi$ of the AP mechanism amounts to $\mathcal{M}(N d_\text{model} B)$ for the dot-product attention of the learned queries $\bm {\bar q}$ and the key chunks $C_l^K$, where the computed attention weights are reused to summarize the value chunks $C_l^V$ as well.
The complexity of PP amounts to $\mathcal{M}(2 (B+1) d_\text{model} d_\text{in} \lceil \frac{N}{M} \rceil)$ for post-processing the attention results of the key and value chunks as described in (\ref{eq:appp}). In order to reduce the computational complexity for the following experiments, the feed-forward neural network of the post-processing stage uses a bottleneck of inner dimension $d_\text{in}=16$.

\vspace{-0.1cm}
\subsection{Related Work}
\vspace{-0.1cm}

The closest prior art to this work is Transformer-XL \cite{DaiYY19} and the recently proposed self-attention with augmented memory \cite{WuWS20}. However, both are different in various ways, e.g., Transformer-XL incorporates history information at each layer only from the previous chunk and from a lower layer. Self-attention with augmented memory uses mean-pooling on chunks of the input sequence to form a summarization query for computing attention weights over the current chunk plus the summarization outputs from previous chunks.
This is a recursive operation, which cannot be easily parallelized, unlike our proposed solution.
Moreover, we use learned embedding vectors for the summarization queries instead of mean-pooling, whereby the characteristics of the relevant frames are learned and which also allows us to use multiple queries per chunk.

\vspace{-0.1cm}
\section{Experiments}
\label{sec:experiments}
\vspace{-0.1cm}

\vspace{-0.1cm}
\subsection{Datasets}
\label{ssec:data}
\vspace{-0.1cm}

The LibriSpeech corpus and the Wall Street Journal (WSJ) corpus are used in this work \cite{PanayotovCPK15,wsj1}. LibriSpeech is a corpus of read English audio books with about 960 hours of training data, 10.7 hours of development data, and 10.5 hours of test data. The development and test data sets are both split into approximately two halves named ``clean'' and ``other'', based on the quality of the recorded speech utterances as assessed using an ASR system \cite{PanayotovCPK15}.
WSJ is a data set of read English newspapers with approximately 81~hours of training data, 1.1~hours of development data, and 0.7~hours of test data \cite{wsj1}.
The average duration of a LibriSpeech utterance amounts to 12.3~seconds, the median duration to 13.8~seconds, and the maximum duration to 29.7~seconds. For the WSJ corpus, the average duration of an utterance is 7.8~seconds, the median duration is 7.7~seconds, and the maximum duration 24.4~seconds.

\vspace{-0.1cm}
\subsection{Settings}
\label{ssec:settings}
\vspace{-0.1cm}

For LibriSpeech-based experiments, the transformer model parameters are $d_\mathrm{model}=512$, $d_\mathrm{ff}=2048$, $d_h=8$, $E=12$, and $D=6$. They are the same for WSJ except $d_\mathrm{model}=256$ and $d_h=4$. The Adam optimizer with $\beta_1=0.9$, $\beta_2=0.98$, $\epsilon=10^{-9}$ and learning rate scheduling similar to \cite{VaswaniSPUJGKP17} is applied for training using 25000 warmup steps. The learning rate factor and the maximum number of training epochs are set to 5.0 and 120 for LibriSpeech, and to 10.0 and 100 for WSJ. 
Output tokens for the LibriSpeech model consist of 5000 subwords obtained by the SentencePiece method \cite{KudoR18}. For WSJ, the number of output tokens amounts to 50, the number of characters in WSJ. Weight factor $\gamma$, which is used to balance the CTC and transformer model objectives during training, is set to 0.3. Layer normalization is applied before and dropout with a rate of 10\% after each $\text{MHA}$ and $\text{FF}$ layer. In addition, label smoothing with a penalty of 0.1 is used and SpecAugment is applied for the LibriSpeech-based ASR experiments \cite{ParkCZC19}.
An RNN-based language model (LM) is employed via shallow fusion for each data sets: a word-based LM of 65k words is applied for WSJ \cite{HoriCW18} and a subword-based LM is applied for LibriSpeech \cite{HoriWZC17}.
The LM weight, CTC weight, and beam size for joint CTC-attention decoding are set to 0.6, 0.4, and 30 for LibriSpeech and to 1.0, 0.3, and 10 for WSJ.

\vspace{-0.1cm}
\subsection{Results}
\label{ssec:results}
\vspace{-0.1cm}

\begin{table}[tb]
 \begin{center}
  \setlength{\aboverulesep}{0pt}
\setlength{\belowrulesep}{0pt}
  \caption{ WSJ-based ASR results for self-attention, restricted self-attention, and dilated self-attention. $R$ denotes the symmetric restriction window size and $M$ the chunk size in number of frames. $\mathcal{M}$ denotes the number of multiplications for self-attention with $N=195$ (average input length) and $d_\text{model}=256$. Numbers after a hyphen denote the number of attention heads for AP. }
  \label{tab:results_wsj}
  \resizebox{0.95\linewidth}{!}
  {%
  \begin{tabular}{lcccccc}
  \specialrule{\heavyrulewidth}{0ex}{0.65ex}
  \multirow{2}{*}[-.05cm]{\shortstack[l]{self-attention\\type}} & \multirow{2}{*}[-.05cm]{\shortstack[c]{dilation\\mechanism}}  &  &  &  & \multicolumn{2}{c}{WER [\%]} \\
  \cmidrule(lr){6-7}
   &  & $R$ & $M$ & $\mathcal{M}$ & dev & test \\
  \specialrule{\lightrulewidth}{0.4ex}{0.65ex}
  full-sequence & - & - & - & 9.8M & \textbf{7.7} & \textbf{4.7} \\
[0.3ex]
\hdashline\noalign{\vskip 0.55ex}
  restricted & - & $35$ & - & 1.8M & 8.7 & 5.9 \\
  restricted & - & $21$ & - & 1.1M & 8.7 & 5.6 \\
  restricted & - & $15$   & - & 0.8M & 8.6 & 5.8 \\
  restricted & - & $13$   & - & 0.7M & 9.0 & 5.7 \\
  restricted & - & $11$   & - & 0.6M & 8.9 & 6.2 \\
  dilated & subsampling & $15$ & 10   & 1.8M & 8.0 & 5.2 \\
  dilated & MP & $15$ & 10   & 1.8M & 7.9 & 5.4 \\
  dilated & AP-1 & $15$ & 11 & 1.8M & 7.7 & 4.9 \\
  dilated & AP-2 & $15$ & 12 & 1.8M & 7.7 & 4.9 \\
  dilated & AP-1+PP & $21$ & 20 & 1.8M & 7.7 & 4.9 \\
  dilated & AP-2+PP & $21$ & 27 & 1.8M & 7.6 & 5.1 \\
  
  dilated & subsampling & $11$ & 20 & 1.1M & 8.1 & 5.3 \\
  dilated & MP & $11$ & 20 & 1.1M & 7.7 & 5.2 \\
  dilated & AP-1 & $11$ & 22 & 1.1M & 7.6 & 5.1 \\
  dilated & AP-2 & $11$ & 28 & 1.1M & 7.7 & 5.1 \\
  dilated & AP-1+PP & $9$ & 24 & 1.1M & 8.2 & 5.2 \\
  dilated & AP-2+PP & $9$ & 33 & 1.1M & \textbf{7.6} & \textbf{4.9} \\
  
\specialrule{\heavyrulewidth}{0.4ex}{0ex}
  \end{tabular}}
  \end{center}
  \vspace{-3mm}
\end{table}

\begin{table}[tb]
 \begin{center}
 \setlength{\aboverulesep}{0pt}
\setlength{\belowrulesep}{0pt}
  \caption{LibriSpeech-based WER [\%] for self-attention, restricted self-attention, and dilated self-attention with results for streaming end-to-end ASR at the bottom \cite{MoritzHLR20}. The computational complexity estimate $\mathcal{M}$ is based on using $d_\text{model}=512$ and $N=310$, which is the average encoder sequence length of a LibriSpeech utterance.} %
  \label{tab:results_libri}
  \resizebox{.95\linewidth}{!}
  {\setlength{\tabcolsep}{2pt}\begin{tabular}{lcccccccc}
  \specialrule{\heavyrulewidth}{0ex}{0.65ex}
  \multirow{2}{*}[-.05cm]{\shortstack[l]{self-attention\\type}} & \multirow{2}{*}[-.05cm]{\shortstack[c]{dilation\\mechanism}} &  &  & \multicolumn{1}{c}{} & \multicolumn{2}{c}{dev} & \multicolumn{2}{c}{test} \\
  \cmidrule(lr){6-7}\cmidrule(lr){8-9}
   &    & $R$ & $M$ & \multicolumn{1}{c}{$\mathcal{M}$} & \multicolumn{1}{c}{clean} & other & clean & other \\
  \specialrule{\lightrulewidth}{0.4ex}{0.65ex}
  full-sequence & - & - & - & 52M & \textbf{2.3} & \textbf{5.7} & \textbf{2.6} & \textbf{6.0} \\
[0.25ex]
\hdashline\noalign{\vskip 0.45ex}
  restricted    & - & 41 & - & 6.5M & 2.5 & 6.5 & 2.7 & 6.6 \\
  restricted    & - & 25 & - & 4.0M & 2.4 & 6.5 & 2.6 & 7.0 \\
  restricted    & - & 13 & - & 2.1M & 2.5 & 7.0 & 2.7 & 7.2 \\
  dilated       & subsampling  & 25 & 20 & 6.5M & 2.3 & 6.1 & 2.6 & 6.2 \\
  dilated   & MP  & 25 & 20 & 6.5M & 2.3 & 6.1 & 2.6 & 6.2 \\
  dilated   & AP-1  & 25 & 20 & 6.7M & 2.3 & 5.8 & 2.5 & 6.1 \\ %
  dilated   & AP-2  & 25 & 20 & 6.8M & 2.3 & 6.1 & 2.5 & 6.1 \\ %
  dilated   & AP-1+PP  & 25 & 20 & 7.2M & 2.3 & 5.9 & 2.5 & 6.1 \\  %
  dilated   & AP-2+PP  & 25 & 20 & 7.6M & \textbf{2.2} & \textbf{5.8} & \textbf{2.4} & \textbf{5.9} \\ %
  dilated   & AP-2+PP  & 17 & 19 & 6.6M & 2.2 & 5.8 & 2.5 & 6.0 \\ %
  
  dilated       & subsampling      & 13 & 40 & 3.3M & 2.4 & 6.5 & 2.6 & 6.7 \\
  dilated   & MP  & 13 & 40 & 3.3M & 2.5 & 6.6 & 2.6 & 6.6 \\
  dilated   & AP-1  & 11 & 34 & 3.5M & 2.4 & 6.0 & 2.5 & 6.3 \\
  dilated   & AP-2+PP  & 11 & 50 & 3.5M & 2.2 & 5.9 & 2.5 & 6.2 \\
  \specialrule{\heavyrulewidth}{0.4ex}{0.65ex}
   \multicolumn{9}{c}{Triggered Attention-based Streaming End-to-End ASR \cite{MoritzHLR20}} \\
   \specialrule{\lightrulewidth}{0.4ex}{0.65ex}
   \multirow{2}{*}[-.05cm]{\shortstack[l]{self-attention\\type}} & \multirow{2}{*}[-.05cm]{\shortstack[c]{dilation\\mechanism}} &  &  & \multicolumn{1}{c}{} & \multicolumn{2}{c}{dev} & \multicolumn{2}{c}{test} \\
   \cmidrule(lr){6-7}\cmidrule(lr){8-9}
    &  & $\bm v_\text{lb}$ & $\bm v_\text{la}$ & \multicolumn{1}{c}{$M$} & \multicolumn{1}{c}{clean}  & other & clean & other \\
   \specialrule{\lightrulewidth}{0.4ex}{0.65ex}
   restricted & -       & $\infty$ & 1 & - & 2.8 & 7.5 & 3.1 & 8.1 \\
  dilated    & AP-2+PP & 9 & 1 & 15 & 2.9 & 7.9 & 3.0 & 8.1 \\
 \specialrule{\heavyrulewidth}{0.4ex}{0ex}
  \end{tabular}
  }
  \end{center}
\end{table}

Table~\ref{tab:results_wsj} shows the WSJ-based ASR results for full-sequence based self-attention, restricted self-attention, and dilated self-attention. Different settings for restricted and dilated self-attention are compared. In addition, subsampling, mean-pooling (MP), and attention-based pooling (AP) with and without post-processing (PP) are investigated, where the number of attention heads $B$ is shown by AP-$B$.
The corresponding computational costs are given as an example for a WSJ utterance of average length (7.8~sec.), which corresponds to a sequence length of 195 frames for a frame rate of 40~ms.
It can be noticed that restricted self-attention considerably increases WERs by more than $1\%$ for all tested restriction window sizes. Dilated self-attention can compensate to a large extent this loss in accuracy, achieving consistently better WERs compared to restricted self-attention.
The overall best performing dilation mechanism is AP, achieving similar results compared to full-sequence based self-attention with more than $88\%$ fewer computational operations.

LibriSpeech ASR results are shown in Table~\ref{tab:results_libri}. Simply using restricted self-attention increases WERs on test-other by $0.6\%$ ($R$=41), $1.0\%$ ($R$=25), and $1.2\%$ ($R$=13). Dilated self-attention almost completely equalizes the disadvantages of restricted self-attention, where the AP-2+PP setup shows the best ASR results.
For example, for $R$=25 and $M$=20, the AP-2+PP system achieves a WER of $2.4\%$ and $5.9\%$ for test-clean and test-other, which is even slightly lower compared to WERs of full-sequence based self-attention with only $15\%$ of the computational costs for self-attention.

Finally, the AP-2+PP approach is applied to a triggered attention (TA) based streaming end-to-end ASR system \cite{MoritzHLR20}, where a TA look-ahead of 12 frames and an encoder look-ahead of 1 frame are used, which results in an algorithmic delay of $480$~ms for the TA-based decoder plus $480$~ms for the encoder.
In this setup, the dilation mechanism only processes past encoder frames, i.e., input frames up to the query frame.
Results demonstrate that dilated self-attention, where the dilation mechanism is applied to past frames only, achieves about similar ASR results compared to using the full past for self-attention, while the computational complexity is substantially reduced.
For example, for computing the next self-attention output with 4 or 8 seconds of past audio input, the computational complexity of self-attention is reduced by a factor 7.2 or 11.8 if no new dilation output is computed and by a factor 1.25 or 2.4 when a new full chunk is processed for extending the dilation sequences $\Delta_i^{[V,K]}$, which is computed every $M$=15 frames.

\vspace{-0.2cm}
\section{Conclusions}
\vspace{-0.1cm}

We proposed a dilation mechanism to reduce the computational complexity of self-attention by enabling attention to an input sequence with different levels of resolution. 
Subsampling, mean-pooling, and attention-based pooling techniques are investigated for summarizing and capturing long-context information in self-attention and to reduce the frame rate of distant information relative to a query.
Dilated self-attention is non-recursive, which allows it to be parallelized for efficient implementation, and it has demonstrated substantial improvements compared to restricted self-attention with attention-based pooling leading to the best ASR results. Dilated self-attention has reached WERs of $2.4\%$ and $5.9\%$ for the test-clean and test-other conditions of LibriSpeech, which is even slightly lower %
than the standard self-attention mode with only $15\%$ of the computational costs for self-attention. In addition, dilated self-attention has been applied to a triggered attention-based ASR system, where it has demonstrated to be effective for streaming ASR as well.

\bibliographystyle{IEEEtran}

\bibliography{refs}

\end{document}